\documentclass[aps,prl,groupedaddress,twocolumn,amsmath,ambsymb]{revtex4-2}

\usepackage{graphicx}
\usepackage{color,colortbl}
\usepackage{xcolor}
\usepackage{soul}

\newcommand{\Rs}{R_{\rm s}}
\newcommand{\Rm}{R_{\rm m}}
\newcommand{\mus}{\mu_{\rm s}}
\newcommand{\mum}{\mu_{\rm m}}
\newcommand{\lams}{\lambda_{\rm s}}
\newcommand{\lamm}{\lambda_{\rm m}}

\newcommand{\eps}{\epsilon}
\newcommand{\epsS}{\epsilon^*_{\rm S}}
\newcommand{\epsM}{\epsilon^*_{\rm M}}
\newcommand{\epsD}{\epsilon^*_{\rm D}}
\newcommand{\SigmaS}{\Sigma^*_{\rm S}}
\newcommand{\SigmaM}{\Sigma^*_{\rm M}}
\newcommand{\SigmaD}{\Sigma^*_{\rm D}}
\newcommand{\GD}{G_{\rm D}}
\newcommand{\GS}{G_{\rm S}}
\newcommand{\GM}{G_{\rm M}}

\begin{document}

\title{Toughness of double network hydrogels: the role of reduced stress propagation}

\author{Samuel B. Walker}
\author{Suzanne M. Fielding}%
\affiliation{Department of Physics, Durham University, Science Laboratories, South Road, Durham DH1 3LE, United Kingdom.}%

\date{\today}

\begin{abstract}
Double network hydrogels show remarkable mechanical performance, combining high strength and fracture toughness with sufficient stiffness to bear load, despite containing only a low density of cross-linked polymer molecules in water. We introduce a simple mesoscale model of a double network material, detailed enough to resolve the salient microphysics of local plastic bond breakage, yet simple enough to address macroscopic cracking.  Load sharing between the networks results in a delocalisation of stress such that the double network  inherits both the stiffness of its stiff-and-brittle sacrificial network and the ductility of its soft-and-ductile matrix network. The underlying mechanism is a reduction in the Eshelby stress propagator between sacrificial bonds, inhibiting the tendency for the plastic failure of one sacrificial bond to propagate stress to neighbouring sacrificial bonds and cause a follow-on cascade of breakages. The mechanism of brittle macroscopic cracking is thereby suppressed, giving instead ductile deformation via diffusely distributed microcracking.  
\end{abstract}

\maketitle

Soft materials are central to numerous important technologies, including biomedical tissue engineering~\cite{lee2001hydrogels}, stretchable electronics~\cite{rogers2010materials}  and soft robotics~\cite{martinez2014soft}. Such applications typically demand large material deformations, making stretchable polymer networks such as hydrogels and elastomers promising candidates. However, these materials often also exhibit brittle failure via crack propagation under only modest loads,  limiting their use in practice. A key technological challenge is to design soft materials that combine high strength and fracture toughness  with sufficient mechanical stiffness  to bear load. 
 
This combination can be achieved in composite materials based on a {\em double} network structure, including double network hydrogels~\cite{gong2010double}, elastomers~\cite{ducrot2014toughening,millereau2018mechanics}, and macroscopic composites~\cite{king2019macroscale}. In pioneering work, Gong et al.~\cite{gong2003double} combined a stiff and brittle ``sacrificial" network in low molar ratio with a soft and ductile ``matrix" network. The resulting
double network hydrogel was  both stiff and ductile.  Indeed, such hydrogels  can achieve remarkable mechanical performance, combining hardness (elastic modulus $0.1-1.0$MPa), strength (failure stress $1-10$MPa, failure strain $10-20$) and  toughness (tearing energy $100-1000$Jm$^{-2}$), independent of deformation rate, and despite comprising up to $90\%$  water~\cite{gong2010double,tanaka2005determination,sun2012highly}. This makes them promising candidates for artificial tendons, ligaments and cartilage~\cite{yasuda2009novel}.   Indeed, many biological networks are themselves multicomponent, including  the cell cytoskeleton~\cite{pollard2018overview}  and extracellular matrix~\cite{frantz2010extracellular}. Biomimetic composites have also been shown to display increased stiffness and strength in vitro~\cite{burla2019stress,jensen2014emergent}.
 
The remarkable toughness of double network materials remains poorly understood theoretically. Lacking in particular is a physical understanding of how a double network structure inhibits the propagation of a macroscopic crack of the kind that would cause catastrophic brittle failure in a single network.  Molecular simulations provide important insights at the microscopic level~\cite{mugnai2024inter,jang2007mechanical,wang2017simulational,higuchi2018fracture,tauber2021sharing}, but struggle to access the system sizes needed to address macroscopic crack propagation (or inhibition).  Continuum models naturally address macroscopic scales~\cite{brown2007model,tanaka2007local,okumura2004toughness,bacca2017model,vernerey2018statistical,lavoie2019continuum,okumura2004toughness,zhao2012theory,wang2011pseudo}, but do not capture the detailed underlying microphysics.

Recent experiments have provided valuable insights. Probing the length scale of internal nonaffine deformations by small angle neutron scattering suggests that stress is delocalised in a double network~\cite{ducrot2015structure}. Birefringent imaging shows that non-optimised double networks have localised stresses whereas optimised ones have more uniformly distributed stress~\cite{gong2003double}. Dynamic light scattering shows the fluctuations of the two networks to be strongly coupled, indicating a mutual entanglement and suggesting that they can share load~\cite{huang2007importance}. A higher molar ratio of matrix network predisposes the creation of multiple microcracks (ductile behaviour) instead of the propagation of single macroscopic crack (brittle failure) ~\cite{fukao2020effect}.  Multi-network elastomers  have wide-spread molecular damage and delocalized strain concentration near a crack, compared with single network materials~\cite{ju2024role}.

Here we introduce a simple mesoscale model of a double network material, detailed enough to resolve the salient microphysics, yet simple enough to address macroscopic cracking. To understand its predictions, which capture all the experimental observations described above, we build on recent progress elucidating  yielding in amorphous materials~\cite{divoux2024ductile,popovic2018elastoplastic,barlow2020ductile,rossi2022finite}, in which a  plastic relaxation event localised in one part of a sample causes an elastic Eshelby  stress propagation to nearby regions~\cite{picard2004elastic}, triggering a follow-on cascade of plastic events and macroscopic cracking. We show how that physics is modified in a double network, with the effective Eshelby propagator between the weak but stiff sacrificial bonds  reduced by load-sharing with the matrix network, for sufficiently high fraction of matrix. The plastic failure of a  sacrificial bond then no longer causes nearby sacrificial bonds to break. This inhibits stress localisation and macroscopic cracking, favouring instead ductile deformation via the formation of multiple diffusely distributed microcracks. The double network then inherits both the stiffness of its stiff but weak sacrificial component and the increased failure strain of its soft but tough matrix.

\begin{figure}[!t]
\includegraphics[width=0.75\columnwidth]{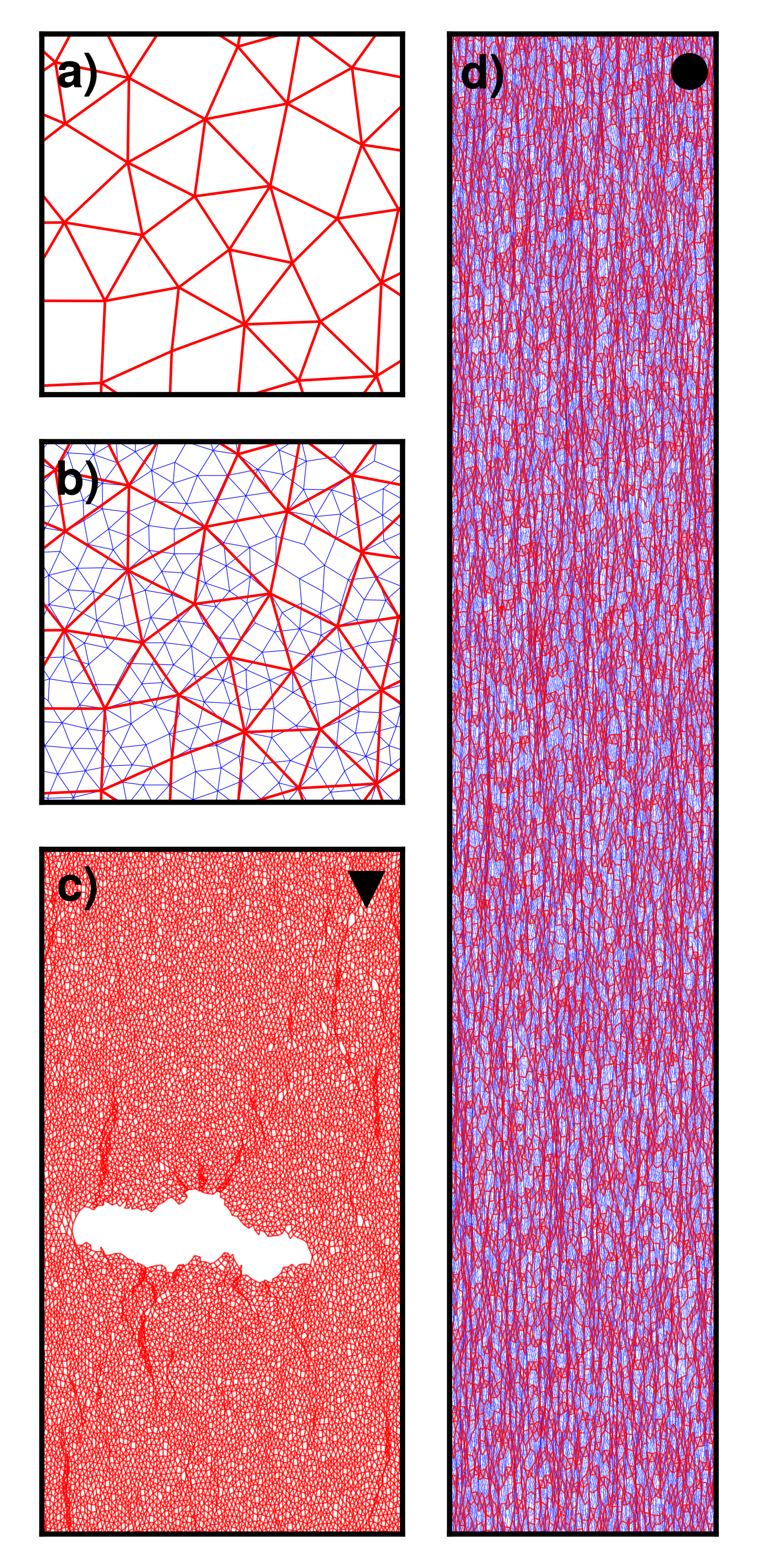}%
\caption{Portion of an {\bf a)} single  and {\bf b)}  double network, with sacrificial bonds in red and matrix bonds  blue. {\bf c)} Stretched single network at a location on the stress-strain curve shown by the triangle in Fig.~\ref{fig:stressStrain}a). A Macroscopic crack has propagated across the network, causing catastrophic material failure. {\bf d)} Stretched double network  at a location on the stress-strain curve shown by the circle in Fig.~\ref{fig:stressStrain}b). Damage has arisen diffusely via many microcracks in the sacrificial network, which do not propagate macroscopically. This allows the double network to stretch much further than the single network without failing.}
\label{fig:networks}
\end{figure}

{\it Model ---} A random double network is prepared by adapting a method commonly used to prepare single networks~\cite{reid2018auxetic}. First, we randomly initialise an ensemble of $N_{\rm S}$ disks of bidisperse radii $\Rs$ and $1.4\Rs$ in equal numbers at area fraction $\phi=1.0$ in a square box, with periodic boundary conditions. With a soft Hookean repulsion between any pair of overlapping disks, the packing is  evolved to equilibrium. Each overlapping pair  then has their centres connected by a spring of stiffness $\mus$ and breakage strain $\lams$, with length $l$ initialised to its equilibrium length $l_0$. The disks are then removed to leave a network of springs. This forms our sacrificial network, Fig.~\ref{fig:networks}a). The matrix network is created likewise, but now starting from $N_{\rm M}$ disks with radii $\Rm$ and $1.4\Rm$, and insisting that one disk of this new ensemble is pinioned to each node of the sacrificial network during equilibration, so as to become a node common to both networks once the disks are then replaced with matrix network springs of stiffness $\mum$ and breakage strain $\lamm$.  We take $M\equiv \Rs/\Rm>1$, so that the sacrificial network (red) is sparse compared with the matrix (blue).  This leaves finally  a double network in which each sacrificial node overlaps with a matrix node, Fig.~\ref{fig:networks}b). Their respective network coordination is $z_{\rm s}=5.42$ (sacrificial) and $z_{\rm m}=5.33$ (matrix).  The nodes common to both networks may represent bonds or entanglement points between them. Either results in load sharing, as seen experimentally~\cite{ahmed2014brittle}. 

\begin{figure}[!t]
\includegraphics[width=0.9\columnwidth]{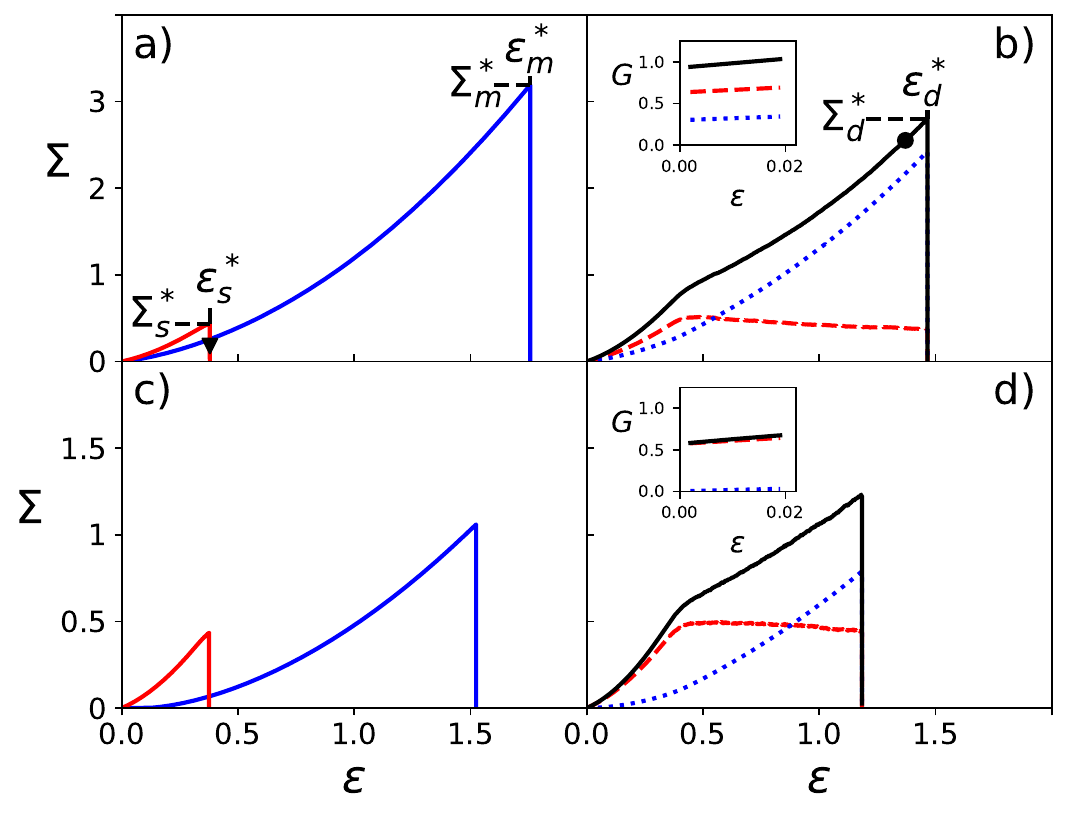}%
\caption{Stress-strain curves. {\bf a)} Single sacrificial network (red solid  line) and  single matrix network (blue solid line). {\bf b)} Composite double network (black solid line), with the contributions of the component sacrificial network (red dashed line) and matrix network (blue dotted line) shown separately.  Inset: modulus $G=d\Sigma/d\epsilon$ at small strain. Triangle and circle show the location of the snapshots in Fig.~\ref{fig:networks}. Parameters:  $M=3.0,  \mum=0.2, \lamm=3.0$. {\bf c)} and {\bf d)} show the effect of reducing the matrix network connectivity to $z_{\rm m}=3.5$, as discussed in the penultimate paragraph of the main text.
}
\label{fig:stressStrain}
\end{figure}

The double network is then stretched  along the $y$ direction at constant area $L_xL_y$  with biperiodic boundary conditions, starting from a square of side $L_{x0}=L_{y0}$. See Fig.~\ref{fig:networks}c,d). The extensional strain $\eps\equiv L_y/L_{y0}-1$. Any bond  breaks immediately once its own strain $l/l_0-1$ exceeds a breakage threshold, $\lams$ or $\lamm$ for sacrificial or matrix bonds respectively.  The stretching proceeds in strain steps of size $\Delta\eps=10^{-3}$ by default, unless more than one spring would immediately break after such a step, in which case we iterate $\Delta\eps$ to find its value,  $\pm 10^{-8}$, that ensures just one bond is (initially) unstable to breakage.  We then allow any ensuing cascade of bond breakages to run to completion, and for the system to equilibrate after it, with r.m.s.  node force $<10^{-6}$, before moving to the next step. We thereby implement quasistatic stretching. The  component $\Sigma_{yy}\equiv\Sigma$ of the network's stress response is calculated using the Irving-Kirkwood formula~\cite{irving1950statistical,hardy1982formulas}.   

\begin{figure}[!t]
\includegraphics[width=0.9\columnwidth]{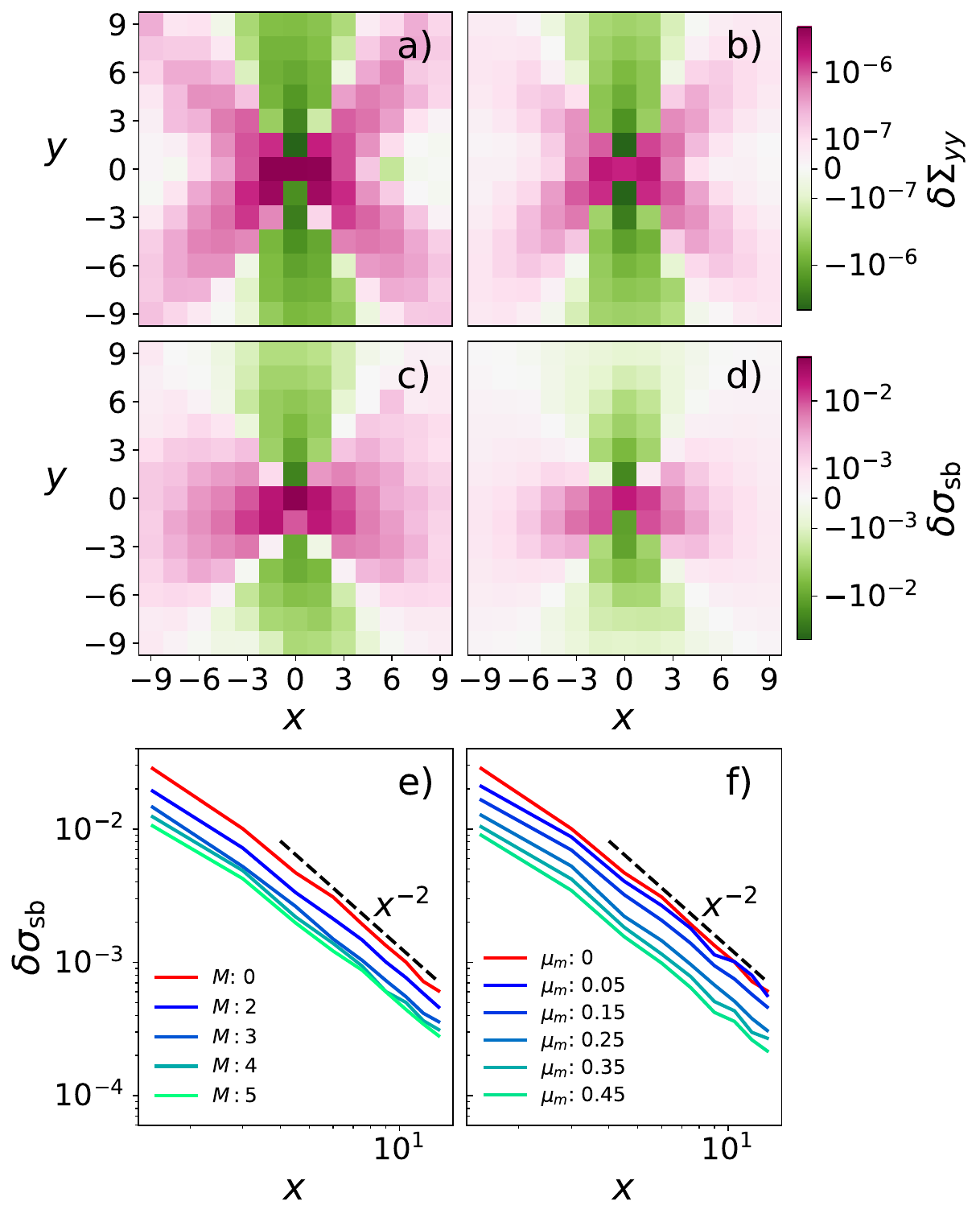}
\caption{Stress propagation to neighbouring sacrificial bonds from the breakage of a sacrificial bond at the origin: (a+c) in a single sacrificial network  and (b+d) in a double network with $M=5.0, \mum=0.2$.  Colourscales show the change due to stress propagation of (a+b)  the $yy$ component  of the Kirkwood stress  $\delta\Sigma_{yy}$; and (c+d)  the sacrificial bond stress  $\delta\sigma_{\rm sb}$, with $\sigma_{\rm sb}\equiv \mus(l-l_0)/l_0$. Each quantity is averaged over all bonds at each grid square. e+f) show the change in bond stress as a function of distance along $x$  at $y=0$ for e) several $M$ at fixed $\mum=0.2$, and f) several $\mum$ at fixed $M=3.0$.  
Lengths are expressed in units such that $\Delta x=1$ corresponds to the typical sacrificial bond length. The propagator of a single sacrificial network is recovered for $M=0$ or $\mum = 0$. 
}
\label{fig:propagator}
\end{figure}

We choose our stress unit $\mus=1.0$, and our length unit as the typical sacrificial bond length, by setting $L_{x0}=L_{y0}=\sqrt{N_{\rm S}}$. We set our  strain  scale $\lams=0.5$ and define the typical number of matrix bond lengths per sacrificial bond length $M=\Rs/\Rm$. Larger $M$ corresponds to a sparser sacrificial network, with more matrix bonds in between. A single sacrificial network is defined to have $M=0$. 
Consistent with the experimental literature, we explore parameters for which, in comparison with the matrix, the sacrificial network is sparse ($M>1.0$), stiff ($\mum<1.0$) and brittle ($\lamm>0.5$).  We take our system size $N_{\rm M}=320^2$, which also specifies $N_{\rm S}$ via $M$.

Fracturing has been studied previously in network models of single component materials~\cite{dussi2020athermal}. A double network model was studied in Ref.~\cite{tauber2020microscopic}, but initialising both a sacrificial bond and a matrix bond between each pair of neighbouring nodes on a triangular lattice. Accordingly, it did not capture the reduction in Eshelby stress propagation as a function of distance between sparse sacrificial bonds with increasing $M$, which is the key physics reported in what follows. Fracture in a double network model with a sparse concentration of unbreakable springs, opposite to the case in real double network hydrogels, has also been studied~\cite{noguchi2024fracture}.

{\it Results ---} Our key finding is summarised as follows. A single network fails  at only modest strain via spatially localised macroscopic cracking, Fig.~\ref{fig:networks}c). In contrast, a double network can be stretched to much larger strain without globally failing. Instead, sacrificial bonds break in numerous microcracks that arise diffusely throughout the network,  Fig.~\ref{fig:networks}d). This distinction is reflected in the corresponding stress-strain curves: Fig.~\ref{fig:stressStrain}a)  for the single network shows a sharp stress drop at only modest overall strain (red curve), whereas Fig.~\ref{fig:stressStrain}d) for the double network shows greatly increased toughness (black curve). 

This can be understood as follows.  In a single network stretched along $y$, the stress lost locally when one bond breaks is propagated  to neighbouring bonds along $x$, putting them under increased stress, Fig.~\ref{fig:propagator}a,c).  These neighbouring bonds then also break, propagating stress to their next neighbours in turn. The resulting knock-on cascade causes  a crack to spread macroscopically along $x$ (Fig.~\ref{fig:networks}c). In contrast,  when a sacrificial bond breaks in a double network,  its stress is partly absorbed by load sharing with nearby matrix bonds, which are tough enough to resist failure. Stress propagation to neighbouring sacrificial bonds is thereby reduced, Fig.~\ref{fig:propagator}b,d). In a well designed network, with model parameter values optimised as described below, this reduction is sufficient to avoid those neighbours then also breaking,  eliminating the knock-on cascade of sacrificial bond breakage  seen during macroscopic cracking in a single network.

We now explore this more quantitatively. As discussed above, a well designed double network has two key properties. (i) It inherits the stiffness of its stiff and brittle sacrificial component, with modulus  $\GD\approx \GS\gg \GM$.
($\GS,\GM,\GD$  are defined via the initial stress-strain slopes in Fig.~\ref{fig:stressStrain}b.)  (ii) It inherits the ductility of its soft and ductile matrix component, with a failure strain $\epsD\approx \epsM\gg \epsS$. (Failure strains and stresses are also defined in Fig.~\ref{fig:stressStrain}a,b).) Our basic hypothesis is that (i) and (ii) stem from the presence of sufficient matrix to reduce the  stress  propagated  between  neighbouring sacrificial bonds; but sufficient sacrificial network to confer high stiffness.  We now explore  parameter values $M$, $\mum$ and $\lamm$ that achieve both (i) and (ii).  

Fig.~\ref{fig:propagator} shows that stress propagation between the stiff but breakable sacrificial network bonds indeed decreases in a double network as the number of  soft but strong matrix bonds  between them, set by $M$, increases at fixed matrix bond stiffness $\mum$; or as $\mum$ increases at fixed $M$. This is to be expected: part of the stress is instead absorbed by the matrix, to an extent that increases with increasing amount of matrix $M$ or increasing matrix stiffness $\mum$. (Conversely, for $M=0$ or $\mum=0$  the level of propagation between neighbouring sacrificial bonds reverts to that in a single network.) These soft matrix bonds can stretch significantly without breaking. In this way, the breakage of a single sacrificial  bond at an origin site has a reduced tendency to cause knock-on breakages in neighbouring sacrificial bonds.

\begin{figure}[!t]
\includegraphics[width=0.95\columnwidth]{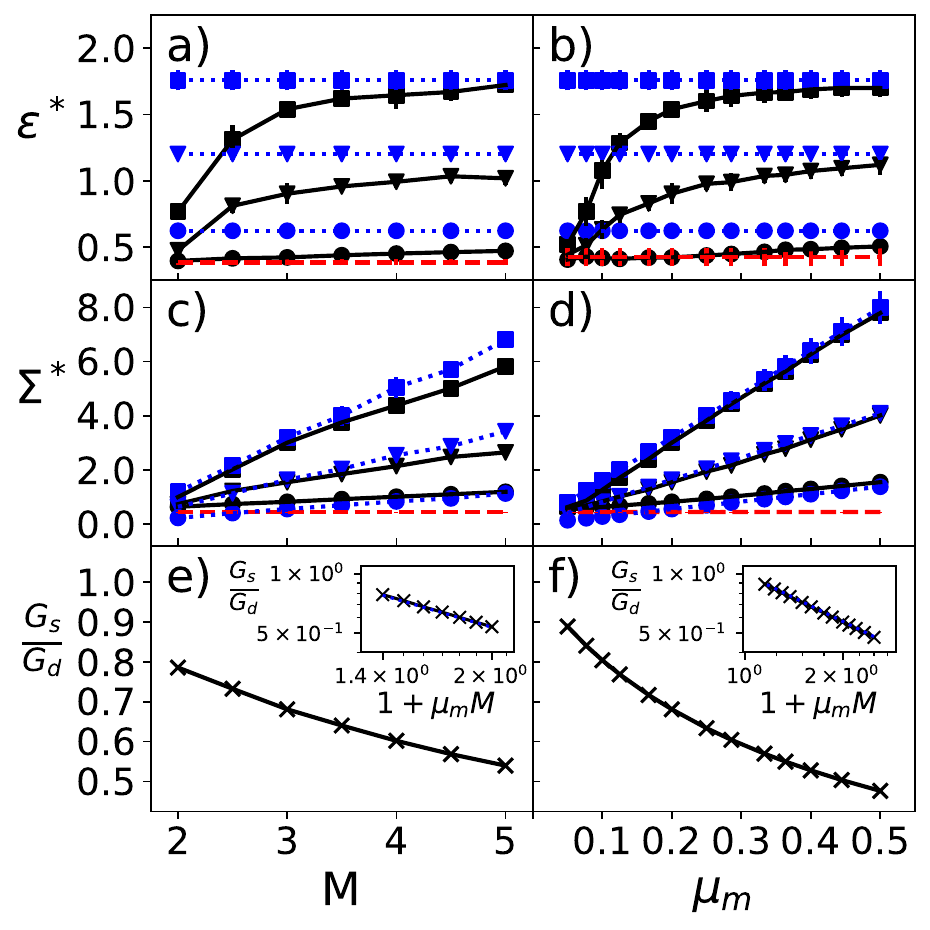}
\caption{{\bf Top:} failure strain of double network (black solid lines) as a function of a) the amount of matrix $M$  and b) matrix bond stiffness $\mum$, shown in each case for matrix bond breakage thresholds $\lamm=1.0, 2.0, 3.0$ in curves bottom to top, marked by circles, triangles and squares respectively.  Failure strains of the corresponding single sacrificial network and single matrix network are shown by red dashed and blue dotted lines respectively. {\bf Middle:} as in a+b) but now showing failure stresses. {\bf Bottom:} modulus of the sacrificial network relative to that of the double network. Matrix bond stiffness $\mum=0.2$ in a,c,e). Amount of matrix $M=3.0$ in b,d,f).}
\label{fig:quantitative}
\end{figure}

As a result of this reduced stress propagation, we expect a double network to be tougher than a single network, with a failure strain $\epsD$ and stress $\SigmaD$ that increases with $M$ or $\mum$. This is confirmed in Figs.~\ref{fig:quantitative}a-d). Indeed, as $M\to\infty$ or $\mum\to\infty$, a double network fully inherits the toughness of the matrix, with failure strain $\epsD\to\epsM$, achieving property (ii) above.  This limit is however trivial: the double network effectively just {\em becomes} a single matrix network in that limit, inheriting  also its softness, so failing to achieve (i). Figs.~\ref{fig:quantitative}e+f) confirm this:  the  sacrificial network contributes progressively less to the modulus $\GD$ of the double network as  $M$ or  $\mum$ increases.

Designing a double network that inherits both the stiffness of its stiff but brittle sacrificial component and the toughness of its soft but ductile matrix component therefore requires a trade-off. In practice, we have found the region of parameter space exemplified by the values $M=3.0,  \mum=0.2, \lamm=3.0$ in Figs.~\ref{fig:networks} and~\ref{fig:stressStrain} to be a sensible compromise.   The single sacrificial network is stiffer than the single matrix network, $\GS > \GM$, but also more brittle, $\epsS < \epsM$ (Fig.~\ref{fig:stressStrain}a). Combining them then gives a double network with a stiffness comparable to that of the sacrificial component and significantly greater than that of the matrix, $\GD=1.47\GS=3.11\GM$ and a toughness comparable to that of  the matrix component, $\epsD\approx \epsM\gg\epsS$, Fig.~\ref{fig:stressStrain}b). Its failure stress $\SigmaD$ greatly exceeds that of the sacrificial network $\SigmaS$ and is of the order of that of the matrix, $\SigmaM$, consistent with a fraction $f=0.41$ of sacrificial bonds having broken diffusely across the material by the time the double network fails, with the sacrificial network no longer percolating.

The model studied here is clearly oversimplified with respect to real double network hydrogels. Indeed, we have taken a $d=2$ dimensional approach and considered bonds with only central forces, without any bending penalty or pre-stretch, in networks with average coordination $z>2d=4$ to ensure rigidity according to Maxwell's criterion. Real gels inhabit $d=3$ dimensions and typically have subcritical coordination numbers $z<2d=6$, with a higher density of cross linkers in the stiff, strongly crosslinked sacrificial network compared with the soft, weakly cross linked matrix network. Furthermore the sacrificial network, which is synthesized first, is naturally pre-stretched by the synthesis of the second matrix within it. We have instead used a single stiffness parameter $\mu$ for each network, with $\mus>\mum$, as a proxy for the greater stiffness of the sacrificial network due to its prestretch and denser crosslinking. This simplified approach has enabled us to identify the key physics at play with the most minimal ingredients possible. 

A first step to exploring these more detailed features is taken in Fig.~\ref{fig:stressStrain}c,d). Here the matrix network's initial connectivity is reduced to $z_{\rm m}=3.5$, with other parameters as before. The double network is still much tougher, with a greatly increased failure strain compared with the single sacrificial network.  However  its stiffness (modulus $G$) is now overwhelmingly dominated by that of the sacrificial matrix, as seen experimentally, due to the now subcritical connectivity of the matrix $z_{\rm m}<4.0$. 

In summary, we have proposed a simple model of double network hydrogels, and shown by numerical simulation that it captures the  key features observed experimentally in these materials. Load sharing between the two networks leads to a delocalisation of stress, such that the double network  inherits both the stiffness of its stiff and brittle sacrificial component and the ductility of its soft and ductile matrix component.  The underlying mechanism is a reduction in the Eshelby stress propagator between neighbouring sacrificial bonds. This inhibits the tendency for the localised plastic failure of a sacrificial bond to propagate stress to neighbouring sacrificial bonds and cause a follow-on cascade of breakages. The mechanism of brittle macroscopic crack formation is thereby suppressed, favouring instead the formation of multiple diffusely distributed microcracks. The trade-off between stiffness and ductility as a function of model parameter values suggests pointers to ongoing material design, both in the protocol explored here and in others such as cyclic shear~\cite{zhang2018fatigue,chen2016improvement,chen2015novel}.

{\it Acknowledgements ---}  This project has received funding from the European Research Council (ERC) under the European Union's Horizon 2020 research and innovation programme (grant agreement No. 885146).

{\it Data availability ---} Data will be made available upon reasonable request.


\begin{thebibliography}{47}%
\makeatletter
\providecommand \@ifxundefined [1]{%
 \@ifx{#1\undefined}
}%
\providecommand \@ifnum [1]{%
 \ifnum #1\expandafter \@firstoftwo
 \else \expandafter \@secondoftwo
 \fi
}%
\providecommand \@ifx [1]{%
 \ifx #1\expandafter \@firstoftwo
 \else \expandafter \@secondoftwo
 \fi
}%
\providecommand \natexlab [1]{#1}%
\providecommand \enquote  [1]{``#1''}%
\providecommand \bibnamefont  [1]{#1}%
\providecommand \bibfnamefont [1]{#1}%
\providecommand \citenamefont [1]{#1}%
\providecommand \href@noop [0]{\@secondoftwo}%
\providecommand \href [0]{\begingroup \@sanitize@url \@href}%
\providecommand \@href[1]{\@@startlink{#1}\@@href}%
\providecommand \@@href[1]{\endgroup#1\@@endlink}%
\providecommand \@sanitize@url [0]{\catcode `\\12\catcode `\$12\catcode `\&12\catcode `\#12\catcode `\^12\catcode `\_12\catcode `\%12\relax}%
\providecommand \@@startlink[1]{}%
\providecommand \@@endlink[0]{}%
\providecommand \url  [0]{\begingroup\@sanitize@url \@url }%
\providecommand \@url [1]{\endgroup\@href {#1}{\urlprefix }}%
\providecommand \urlprefix  [0]{URL }%
\providecommand \Eprint [0]{\href }%
\providecommand \doibase [0]{https://doi.org/}%
\providecommand \selectlanguage [0]{\@gobble}%
\providecommand \bibinfo  [0]{\@secondoftwo}%
\providecommand \bibfield  [0]{\@secondoftwo}%
\providecommand \translation [1]{[#1]}%
\providecommand \BibitemOpen [0]{}%
\providecommand \bibitemStop [0]{}%
\providecommand \bibitemNoStop [0]{.\EOS\space}%
\providecommand \EOS [0]{\spacefactor3000\relax}%
\providecommand \BibitemShut  [1]{\csname bibitem#1\endcsname}%
\let\auto@bib@innerbib\@empty
\bibitem [{\citenamefont {Lee}\ and\ \citenamefont {Mooney}(2001)}]{lee2001hydrogels}%
  \BibitemOpen
  \bibfield  {author} {\bibinfo {author} {\bibfnamefont {K.~Y.}\ \bibnamefont {Lee}}\ and\ \bibinfo {author} {\bibfnamefont {D.~J.}\ \bibnamefont {Mooney}},\ }\bibfield  {title} {\bibinfo {title} {Hydrogels for tissue engineering},\ }\href@noop {} {\bibfield  {journal} {\bibinfo  {journal} {Chemical reviews}\ }\textbf {\bibinfo {volume} {101}},\ \bibinfo {pages} {1869} (\bibinfo {year} {2001})}\BibitemShut {NoStop}%
\bibitem [{\citenamefont {Rogers}\ \emph {et~al.}(2010)\citenamefont {Rogers}, \citenamefont {Someya},\ and\ \citenamefont {Huang}}]{rogers2010materials}%
  \BibitemOpen
  \bibfield  {author} {\bibinfo {author} {\bibfnamefont {J.~A.}\ \bibnamefont {Rogers}}, \bibinfo {author} {\bibfnamefont {T.}~\bibnamefont {Someya}},\ and\ \bibinfo {author} {\bibfnamefont {Y.}~\bibnamefont {Huang}},\ }\bibfield  {title} {\bibinfo {title} {Materials and mechanics for stretchable electronics},\ }\href@noop {} {\bibfield  {journal} {\bibinfo  {journal} {science}\ }\textbf {\bibinfo {volume} {327}},\ \bibinfo {pages} {1603} (\bibinfo {year} {2010})}\BibitemShut {NoStop}%
\bibitem [{\citenamefont {Martinez}\ \emph {et~al.}(2014)\citenamefont {Martinez}, \citenamefont {Glavan}, \citenamefont {Keplinger}, \citenamefont {Oyetibo},\ and\ \citenamefont {Whitesides}}]{martinez2014soft}%
  \BibitemOpen
  \bibfield  {author} {\bibinfo {author} {\bibfnamefont {R.~V.}\ \bibnamefont {Martinez}}, \bibinfo {author} {\bibfnamefont {A.~C.}\ \bibnamefont {Glavan}}, \bibinfo {author} {\bibfnamefont {C.}~\bibnamefont {Keplinger}}, \bibinfo {author} {\bibfnamefont {A.~I.}\ \bibnamefont {Oyetibo}},\ and\ \bibinfo {author} {\bibfnamefont {G.~M.}\ \bibnamefont {Whitesides}},\ }\bibfield  {title} {\bibinfo {title} {Soft actuators and robots that are resistant to mechanical damage},\ }\href@noop {} {\bibfield  {journal} {\bibinfo  {journal} {Advanced Functional Materials}\ }\textbf {\bibinfo {volume} {24}},\ \bibinfo {pages} {3003} (\bibinfo {year} {2014})}\BibitemShut {NoStop}%
\bibitem [{\citenamefont {Gong}(2010)}]{gong2010double}%
  \BibitemOpen
  \bibfield  {author} {\bibinfo {author} {\bibfnamefont {J.~P.}\ \bibnamefont {Gong}},\ }\bibfield  {title} {\bibinfo {title} {Why are double network hydrogels so tough?},\ }\href@noop {} {\bibfield  {journal} {\bibinfo  {journal} {Soft Matter}\ }\textbf {\bibinfo {volume} {6}},\ \bibinfo {pages} {2583} (\bibinfo {year} {2010})}\BibitemShut {NoStop}%
\bibitem [{\citenamefont {Ducrot}\ \emph {et~al.}(2014)\citenamefont {Ducrot}, \citenamefont {Chen}, \citenamefont {Bulters}, \citenamefont {Sijbesma},\ and\ \citenamefont {Creton}}]{ducrot2014toughening}%
  \BibitemOpen
  \bibfield  {author} {\bibinfo {author} {\bibfnamefont {E.}~\bibnamefont {Ducrot}}, \bibinfo {author} {\bibfnamefont {Y.}~\bibnamefont {Chen}}, \bibinfo {author} {\bibfnamefont {M.}~\bibnamefont {Bulters}}, \bibinfo {author} {\bibfnamefont {R.~P.}\ \bibnamefont {Sijbesma}},\ and\ \bibinfo {author} {\bibfnamefont {C.}~\bibnamefont {Creton}},\ }\bibfield  {title} {\bibinfo {title} {Toughening elastomers with sacrificial bonds and watching them break},\ }\href@noop {} {\bibfield  {journal} {\bibinfo  {journal} {Science}\ }\textbf {\bibinfo {volume} {344}},\ \bibinfo {pages} {186} (\bibinfo {year} {2014})}\BibitemShut {NoStop}%
\bibitem [{\citenamefont {Millereau}\ \emph {et~al.}(2018)\citenamefont {Millereau}, \citenamefont {Ducrot}, \citenamefont {Clough}, \citenamefont {Wiseman}, \citenamefont {Brown}, \citenamefont {Sijbesma},\ and\ \citenamefont {Creton}}]{millereau2018mechanics}%
  \BibitemOpen
  \bibfield  {author} {\bibinfo {author} {\bibfnamefont {P.}~\bibnamefont {Millereau}}, \bibinfo {author} {\bibfnamefont {E.}~\bibnamefont {Ducrot}}, \bibinfo {author} {\bibfnamefont {J.~M.}\ \bibnamefont {Clough}}, \bibinfo {author} {\bibfnamefont {M.~E.}\ \bibnamefont {Wiseman}}, \bibinfo {author} {\bibfnamefont {H.~R.}\ \bibnamefont {Brown}}, \bibinfo {author} {\bibfnamefont {R.~P.}\ \bibnamefont {Sijbesma}},\ and\ \bibinfo {author} {\bibfnamefont {C.}~\bibnamefont {Creton}},\ }\bibfield  {title} {\bibinfo {title} {Mechanics of elastomeric molecular composites},\ }\href@noop {} {\bibfield  {journal} {\bibinfo  {journal} {Proceedings of the National Academy of Sciences}\ }\textbf {\bibinfo {volume} {115}},\ \bibinfo {pages} {9110} (\bibinfo {year} {2018})}\BibitemShut {NoStop}%
\bibitem [{\citenamefont {King}\ \emph {et~al.}(2019)\citenamefont {King}, \citenamefont {Okumura}, \citenamefont {Takahashi}, \citenamefont {Kurokawa},\ and\ \citenamefont {Gong}}]{king2019macroscale}%
  \BibitemOpen
  \bibfield  {author} {\bibinfo {author} {\bibfnamefont {D.~R.}\ \bibnamefont {King}}, \bibinfo {author} {\bibfnamefont {T.}~\bibnamefont {Okumura}}, \bibinfo {author} {\bibfnamefont {R.}~\bibnamefont {Takahashi}}, \bibinfo {author} {\bibfnamefont {T.}~\bibnamefont {Kurokawa}},\ and\ \bibinfo {author} {\bibfnamefont {J.~P.}\ \bibnamefont {Gong}},\ }\bibfield  {title} {\bibinfo {title} {Macroscale double networks: design criteria for optimizing strength and toughness},\ }\href@noop {} {\bibfield  {journal} {\bibinfo  {journal} {ACS applied materials \& interfaces}\ }\textbf {\bibinfo {volume} {11}},\ \bibinfo {pages} {35343} (\bibinfo {year} {2019})}\BibitemShut {NoStop}%
\bibitem [{\citenamefont {Gong}\ \emph {et~al.}(2003)\citenamefont {Gong}, \citenamefont {Katsuyama}, \citenamefont {Kurokawa},\ and\ \citenamefont {Osada}}]{gong2003double}%
  \BibitemOpen
  \bibfield  {author} {\bibinfo {author} {\bibfnamefont {J.~P.}\ \bibnamefont {Gong}}, \bibinfo {author} {\bibfnamefont {Y.}~\bibnamefont {Katsuyama}}, \bibinfo {author} {\bibfnamefont {T.}~\bibnamefont {Kurokawa}},\ and\ \bibinfo {author} {\bibfnamefont {Y.}~\bibnamefont {Osada}},\ }\bibfield  {title} {\bibinfo {title} {Double-network hydrogels with extremely high mechanical strength},\ }\href@noop {} {\bibfield  {journal} {\bibinfo  {journal} {Advanced materials}\ }\textbf {\bibinfo {volume} {15}},\ \bibinfo {pages} {1155} (\bibinfo {year} {2003})}\BibitemShut {NoStop}%
\bibitem [{\citenamefont {Tanaka}\ \emph {et~al.}(2005)\citenamefont {Tanaka}, \citenamefont {Kuwabara}, \citenamefont {Na}, \citenamefont {Kurokawa}, \citenamefont {Gong},\ and\ \citenamefont {Osada}}]{tanaka2005determination}%
  \BibitemOpen
  \bibfield  {author} {\bibinfo {author} {\bibfnamefont {Y.}~\bibnamefont {Tanaka}}, \bibinfo {author} {\bibfnamefont {R.}~\bibnamefont {Kuwabara}}, \bibinfo {author} {\bibfnamefont {Y.-H.}\ \bibnamefont {Na}}, \bibinfo {author} {\bibfnamefont {T.}~\bibnamefont {Kurokawa}}, \bibinfo {author} {\bibfnamefont {J.~P.}\ \bibnamefont {Gong}},\ and\ \bibinfo {author} {\bibfnamefont {Y.}~\bibnamefont {Osada}},\ }\bibfield  {title} {\bibinfo {title} {Determination of fracture energy of high strength double network hydrogels},\ }\href@noop {} {\bibfield  {journal} {\bibinfo  {journal} {The Journal of Physical Chemistry B}\ }\textbf {\bibinfo {volume} {109}},\ \bibinfo {pages} {11559} (\bibinfo {year} {2005})}\BibitemShut {NoStop}%
\bibitem [{\citenamefont {Sun}\ \emph {et~al.}(2012)\citenamefont {Sun}, \citenamefont {Zhao}, \citenamefont {Illeperuma}, \citenamefont {Chaudhuri}, \citenamefont {Oh}, \citenamefont {Mooney}, \citenamefont {Vlassak},\ and\ \citenamefont {Suo}}]{sun2012highly}%
  \BibitemOpen
  \bibfield  {author} {\bibinfo {author} {\bibfnamefont {J.-Y.}\ \bibnamefont {Sun}}, \bibinfo {author} {\bibfnamefont {X.}~\bibnamefont {Zhao}}, \bibinfo {author} {\bibfnamefont {W.~R.}\ \bibnamefont {Illeperuma}}, \bibinfo {author} {\bibfnamefont {O.}~\bibnamefont {Chaudhuri}}, \bibinfo {author} {\bibfnamefont {K.~H.}\ \bibnamefont {Oh}}, \bibinfo {author} {\bibfnamefont {D.~J.}\ \bibnamefont {Mooney}}, \bibinfo {author} {\bibfnamefont {J.~J.}\ \bibnamefont {Vlassak}},\ and\ \bibinfo {author} {\bibfnamefont {Z.}~\bibnamefont {Suo}},\ }\bibfield  {title} {\bibinfo {title} {Highly stretchable and tough hydrogels},\ }\href@noop {} {\bibfield  {journal} {\bibinfo  {journal} {Nature}\ }\textbf {\bibinfo {volume} {489}},\ \bibinfo {pages} {133} (\bibinfo {year} {2012})}\BibitemShut {NoStop}%
\bibitem [{\citenamefont {Yasuda}\ \emph {et~al.}(2009)\citenamefont {Yasuda}, \citenamefont {Kitamura}, \citenamefont {Gong}, \citenamefont {Arakaki}, \citenamefont {Kwon}, \citenamefont {Onodera}, \citenamefont {Chen}, \citenamefont {Kurokawa}, \citenamefont {Kanaya}, \citenamefont {Ohmiya} \emph {et~al.}}]{yasuda2009novel}%
  \BibitemOpen
  \bibfield  {author} {\bibinfo {author} {\bibfnamefont {K.}~\bibnamefont {Yasuda}}, \bibinfo {author} {\bibfnamefont {N.}~\bibnamefont {Kitamura}}, \bibinfo {author} {\bibfnamefont {J.~P.}\ \bibnamefont {Gong}}, \bibinfo {author} {\bibfnamefont {K.}~\bibnamefont {Arakaki}}, \bibinfo {author} {\bibfnamefont {H.~J.}\ \bibnamefont {Kwon}}, \bibinfo {author} {\bibfnamefont {S.}~\bibnamefont {Onodera}}, \bibinfo {author} {\bibfnamefont {Y.~M.}\ \bibnamefont {Chen}}, \bibinfo {author} {\bibfnamefont {T.}~\bibnamefont {Kurokawa}}, \bibinfo {author} {\bibfnamefont {F.}~\bibnamefont {Kanaya}}, \bibinfo {author} {\bibfnamefont {Y.}~\bibnamefont {Ohmiya}}, \emph {et~al.},\ }\bibfield  {title} {\bibinfo {title} {A novel double-network hydrogel induces spontaneous articular cartilage regeneration in vivo in a large osteochondral defect},\ }\href@noop {} {\bibfield  {journal} {\bibinfo  {journal} {Macromolecular bioscience}\ }\textbf {\bibinfo {volume} {9}},\ \bibinfo {pages} {307} (\bibinfo {year} {2009})}\BibitemShut
  {NoStop}%
\bibitem [{\citenamefont {Pollard}\ and\ \citenamefont {Goldman}(2018)}]{pollard2018overview}%
  \BibitemOpen
  \bibfield  {author} {\bibinfo {author} {\bibfnamefont {T.~D.}\ \bibnamefont {Pollard}}\ and\ \bibinfo {author} {\bibfnamefont {R.~D.}\ \bibnamefont {Goldman}},\ }\bibfield  {title} {\bibinfo {title} {Overview of the cytoskeleton from an evolutionary perspective},\ }\href@noop {} {\bibfield  {journal} {\bibinfo  {journal} {Cold Spring Harbor perspectives in biology}\ }\textbf {\bibinfo {volume} {10}},\ \bibinfo {pages} {a030288} (\bibinfo {year} {2018})}\BibitemShut {NoStop}%
\bibitem [{\citenamefont {Frantz}\ \emph {et~al.}(2010)\citenamefont {Frantz}, \citenamefont {Stewart},\ and\ \citenamefont {Weaver}}]{frantz2010extracellular}%
  \BibitemOpen
  \bibfield  {author} {\bibinfo {author} {\bibfnamefont {C.}~\bibnamefont {Frantz}}, \bibinfo {author} {\bibfnamefont {K.~M.}\ \bibnamefont {Stewart}},\ and\ \bibinfo {author} {\bibfnamefont {V.~M.}\ \bibnamefont {Weaver}},\ }\bibfield  {title} {\bibinfo {title} {The extracellular matrix at a glance},\ }\href@noop {} {\bibfield  {journal} {\bibinfo  {journal} {Journal of cell science}\ }\textbf {\bibinfo {volume} {123}},\ \bibinfo {pages} {4195} (\bibinfo {year} {2010})}\BibitemShut {NoStop}%
\bibitem [{\citenamefont {Burla}\ \emph {et~al.}(2019)\citenamefont {Burla}, \citenamefont {Tauber}, \citenamefont {Dussi}, \citenamefont {van Der~Gucht},\ and\ \citenamefont {Koenderink}}]{burla2019stress}%
  \BibitemOpen
  \bibfield  {author} {\bibinfo {author} {\bibfnamefont {F.}~\bibnamefont {Burla}}, \bibinfo {author} {\bibfnamefont {J.}~\bibnamefont {Tauber}}, \bibinfo {author} {\bibfnamefont {S.}~\bibnamefont {Dussi}}, \bibinfo {author} {\bibfnamefont {J.}~\bibnamefont {van Der~Gucht}},\ and\ \bibinfo {author} {\bibfnamefont {G.~H.}\ \bibnamefont {Koenderink}},\ }\bibfield  {title} {\bibinfo {title} {Stress management in composite biopolymer networks},\ }\href@noop {} {\bibfield  {journal} {\bibinfo  {journal} {Nature physics}\ }\textbf {\bibinfo {volume} {15}},\ \bibinfo {pages} {549} (\bibinfo {year} {2019})}\BibitemShut {NoStop}%
\bibitem [{\citenamefont {Jensen}\ \emph {et~al.}(2014)\citenamefont {Jensen}, \citenamefont {Morris}, \citenamefont {Goldman},\ and\ \citenamefont {Weitz}}]{jensen2014emergent}%
  \BibitemOpen
  \bibfield  {author} {\bibinfo {author} {\bibfnamefont {M.~H.}\ \bibnamefont {Jensen}}, \bibinfo {author} {\bibfnamefont {E.~J.}\ \bibnamefont {Morris}}, \bibinfo {author} {\bibfnamefont {R.~D.}\ \bibnamefont {Goldman}},\ and\ \bibinfo {author} {\bibfnamefont {D.~A.}\ \bibnamefont {Weitz}},\ }\bibfield  {title} {\bibinfo {title} {Emergent properties of composite semiflexible biopolymer networks},\ }\href@noop {} {\bibfield  {journal} {\bibinfo  {journal} {Bioarchitecture}\ }\textbf {\bibinfo {volume} {4}},\ \bibinfo {pages} {138} (\bibinfo {year} {2014})}\BibitemShut {NoStop}%
\bibitem [{\citenamefont {Mugnai}\ \emph {et~al.}(2024)\citenamefont {Mugnai}, \citenamefont {Batoum},\ and\ \citenamefont {Del~Gado}}]{mugnai2024inter}%
  \BibitemOpen
  \bibfield  {author} {\bibinfo {author} {\bibfnamefont {M.~L.}\ \bibnamefont {Mugnai}}, \bibinfo {author} {\bibfnamefont {R.~T.}\ \bibnamefont {Batoum}},\ and\ \bibinfo {author} {\bibfnamefont {E.}~\bibnamefont {Del~Gado}},\ }\bibfield  {title} {\bibinfo {title} {Inter-species interactions in dual, fibrous gel enable control of gel structure and rheology},\ }\href@noop {} {\bibfield  {journal} {\bibinfo  {journal} {arXiv preprint arXiv:2411.09665}\ } (\bibinfo {year} {2024})}\BibitemShut {NoStop}%
\bibitem [{\citenamefont {Jang}\ \emph {et~al.}(2007)\citenamefont {Jang}, \citenamefont {Goddard},\ and\ \citenamefont {Kalani}}]{jang2007mechanical}%
  \BibitemOpen
  \bibfield  {author} {\bibinfo {author} {\bibfnamefont {S.~S.}\ \bibnamefont {Jang}}, \bibinfo {author} {\bibfnamefont {W.~A.}\ \bibnamefont {Goddard}},\ and\ \bibinfo {author} {\bibfnamefont {M.~Y.~S.}\ \bibnamefont {Kalani}},\ }\bibfield  {title} {\bibinfo {title} {Mechanical and transport properties of the poly (ethylene oxide)- poly (acrylic acid) double network hydrogel from molecular dynamic simulations},\ }\href@noop {} {\bibfield  {journal} {\bibinfo  {journal} {The Journal of Physical Chemistry B}\ }\textbf {\bibinfo {volume} {111}},\ \bibinfo {pages} {1729} (\bibinfo {year} {2007})}\BibitemShut {NoStop}%
\bibitem [{\citenamefont {Wang}\ \emph {et~al.}(2017)\citenamefont {Wang}, \citenamefont {Zhang}, \citenamefont {Davris}, \citenamefont {Liu}, \citenamefont {Gao}, \citenamefont {Zhang},\ and\ \citenamefont {Lyulin}}]{wang2017simulational}%
  \BibitemOpen
  \bibfield  {author} {\bibinfo {author} {\bibfnamefont {W.}~\bibnamefont {Wang}}, \bibinfo {author} {\bibfnamefont {Z.}~\bibnamefont {Zhang}}, \bibinfo {author} {\bibfnamefont {T.}~\bibnamefont {Davris}}, \bibinfo {author} {\bibfnamefont {J.}~\bibnamefont {Liu}}, \bibinfo {author} {\bibfnamefont {Y.}~\bibnamefont {Gao}}, \bibinfo {author} {\bibfnamefont {L.}~\bibnamefont {Zhang}},\ and\ \bibinfo {author} {\bibfnamefont {A.~V.}\ \bibnamefont {Lyulin}},\ }\bibfield  {title} {\bibinfo {title} {Simulational insights into the mechanical response of prestretched double network filled elastomers},\ }\href@noop {} {\bibfield  {journal} {\bibinfo  {journal} {Soft Matter}\ }\textbf {\bibinfo {volume} {13}},\ \bibinfo {pages} {8597} (\bibinfo {year} {2017})}\BibitemShut {NoStop}%
\bibitem [{\citenamefont {Higuchi}\ \emph {et~al.}(2018)\citenamefont {Higuchi}, \citenamefont {Saito}, \citenamefont {Sakai}, \citenamefont {Gong},\ and\ \citenamefont {Kubo}}]{higuchi2018fracture}%
  \BibitemOpen
  \bibfield  {author} {\bibinfo {author} {\bibfnamefont {Y.}~\bibnamefont {Higuchi}}, \bibinfo {author} {\bibfnamefont {K.}~\bibnamefont {Saito}}, \bibinfo {author} {\bibfnamefont {T.}~\bibnamefont {Sakai}}, \bibinfo {author} {\bibfnamefont {J.~P.}\ \bibnamefont {Gong}},\ and\ \bibinfo {author} {\bibfnamefont {M.}~\bibnamefont {Kubo}},\ }\bibfield  {title} {\bibinfo {title} {Fracture process of double-network gels by coarse-grained molecular dynamics simulation},\ }\href@noop {} {\bibfield  {journal} {\bibinfo  {journal} {Macromolecules}\ }\textbf {\bibinfo {volume} {51}},\ \bibinfo {pages} {3075} (\bibinfo {year} {2018})}\BibitemShut {NoStop}%
\bibitem [{\citenamefont {Tauber}\ \emph {et~al.}(2021)\citenamefont {Tauber}, \citenamefont {Rovigatti}, \citenamefont {Dussi},\ and\ \citenamefont {Van Der~Gucht}}]{tauber2021sharing}%
  \BibitemOpen
  \bibfield  {author} {\bibinfo {author} {\bibfnamefont {J.}~\bibnamefont {Tauber}}, \bibinfo {author} {\bibfnamefont {L.}~\bibnamefont {Rovigatti}}, \bibinfo {author} {\bibfnamefont {S.}~\bibnamefont {Dussi}},\ and\ \bibinfo {author} {\bibfnamefont {J.}~\bibnamefont {Van Der~Gucht}},\ }\bibfield  {title} {\bibinfo {title} {Sharing the load: Stress redistribution governs fracture of polymer double networks},\ }\href@noop {} {\bibfield  {journal} {\bibinfo  {journal} {Macromolecules}\ }\textbf {\bibinfo {volume} {54}},\ \bibinfo {pages} {8563} (\bibinfo {year} {2021})}\BibitemShut {NoStop}%
\bibitem [{\citenamefont {Brown}(2007)}]{brown2007model}%
  \BibitemOpen
  \bibfield  {author} {\bibinfo {author} {\bibfnamefont {H.~R.}\ \bibnamefont {Brown}},\ }\bibfield  {title} {\bibinfo {title} {A model of the fracture of double network gels},\ }\href@noop {} {\bibfield  {journal} {\bibinfo  {journal} {Macromolecules}\ }\textbf {\bibinfo {volume} {40}},\ \bibinfo {pages} {3815} (\bibinfo {year} {2007})}\BibitemShut {NoStop}%
\bibitem [{\citenamefont {Tanaka}(2007)}]{tanaka2007local}%
  \BibitemOpen
  \bibfield  {author} {\bibinfo {author} {\bibfnamefont {Y.}~\bibnamefont {Tanaka}},\ }\bibfield  {title} {\bibinfo {title} {A local damage model for anomalous high toughness of double-network gels},\ }\href@noop {} {\bibfield  {journal} {\bibinfo  {journal} {Europhysics Letters}\ }\textbf {\bibinfo {volume} {78}},\ \bibinfo {pages} {56005} (\bibinfo {year} {2007})}\BibitemShut {NoStop}%
\bibitem [{\citenamefont {Okumura}(2004)}]{okumura2004toughness}%
  \BibitemOpen
  \bibfield  {author} {\bibinfo {author} {\bibfnamefont {K.}~\bibnamefont {Okumura}},\ }\bibfield  {title} {\bibinfo {title} {Toughness of double elastic networks},\ }\href@noop {} {\bibfield  {journal} {\bibinfo  {journal} {Europhysics Letters}\ }\textbf {\bibinfo {volume} {67}},\ \bibinfo {pages} {470} (\bibinfo {year} {2004})}\BibitemShut {NoStop}%
\bibitem [{\citenamefont {Bacca}\ \emph {et~al.}(2017)\citenamefont {Bacca}, \citenamefont {Creton},\ and\ \citenamefont {McMeeking}}]{bacca2017model}%
  \BibitemOpen
  \bibfield  {author} {\bibinfo {author} {\bibfnamefont {M.}~\bibnamefont {Bacca}}, \bibinfo {author} {\bibfnamefont {C.}~\bibnamefont {Creton}},\ and\ \bibinfo {author} {\bibfnamefont {R.~M.}\ \bibnamefont {McMeeking}},\ }\bibfield  {title} {\bibinfo {title} {A model for the mullins effect in multinetwork elastomers},\ }\href@noop {} {\bibfield  {journal} {\bibinfo  {journal} {Journal of Applied Mechanics}\ }\textbf {\bibinfo {volume} {84}},\ \bibinfo {pages} {121009} (\bibinfo {year} {2017})}\BibitemShut {NoStop}%
\bibitem [{\citenamefont {Vernerey}\ \emph {et~al.}(2018)\citenamefont {Vernerey}, \citenamefont {Brighenti}, \citenamefont {Long},\ and\ \citenamefont {Shen}}]{vernerey2018statistical}%
  \BibitemOpen
  \bibfield  {author} {\bibinfo {author} {\bibfnamefont {F.~J.}\ \bibnamefont {Vernerey}}, \bibinfo {author} {\bibfnamefont {R.}~\bibnamefont {Brighenti}}, \bibinfo {author} {\bibfnamefont {R.}~\bibnamefont {Long}},\ and\ \bibinfo {author} {\bibfnamefont {T.}~\bibnamefont {Shen}},\ }\bibfield  {title} {\bibinfo {title} {Statistical damage mechanics of polymer networks},\ }\href@noop {} {\bibfield  {journal} {\bibinfo  {journal} {Macromolecules}\ }\textbf {\bibinfo {volume} {51}},\ \bibinfo {pages} {6609} (\bibinfo {year} {2018})}\BibitemShut {NoStop}%
\bibitem [{\citenamefont {Lavoie}\ \emph {et~al.}(2019)\citenamefont {Lavoie}, \citenamefont {Millereau}, \citenamefont {Creton}, \citenamefont {Long},\ and\ \citenamefont {Tang}}]{lavoie2019continuum}%
  \BibitemOpen
  \bibfield  {author} {\bibinfo {author} {\bibfnamefont {S.~R.}\ \bibnamefont {Lavoie}}, \bibinfo {author} {\bibfnamefont {P.}~\bibnamefont {Millereau}}, \bibinfo {author} {\bibfnamefont {C.}~\bibnamefont {Creton}}, \bibinfo {author} {\bibfnamefont {R.}~\bibnamefont {Long}},\ and\ \bibinfo {author} {\bibfnamefont {T.}~\bibnamefont {Tang}},\ }\bibfield  {title} {\bibinfo {title} {A continuum model for progressive damage in tough multinetwork elastomers},\ }\href@noop {} {\bibfield  {journal} {\bibinfo  {journal} {Journal of the Mechanics and Physics of Solids}\ }\textbf {\bibinfo {volume} {125}},\ \bibinfo {pages} {523} (\bibinfo {year} {2019})}\BibitemShut {NoStop}%
\bibitem [{\citenamefont {Zhao}(2012)}]{zhao2012theory}%
  \BibitemOpen
  \bibfield  {author} {\bibinfo {author} {\bibfnamefont {X.}~\bibnamefont {Zhao}},\ }\bibfield  {title} {\bibinfo {title} {A theory for large deformation and damage of interpenetrating polymer networks},\ }\href@noop {} {\bibfield  {journal} {\bibinfo  {journal} {Journal of the Mechanics and Physics of Solids}\ }\textbf {\bibinfo {volume} {60}},\ \bibinfo {pages} {319} (\bibinfo {year} {2012})}\BibitemShut {NoStop}%
\bibitem [{\citenamefont {Wang}\ and\ \citenamefont {Hong}(2011)}]{wang2011pseudo}%
  \BibitemOpen
  \bibfield  {author} {\bibinfo {author} {\bibfnamefont {X.}~\bibnamefont {Wang}}\ and\ \bibinfo {author} {\bibfnamefont {W.}~\bibnamefont {Hong}},\ }\bibfield  {title} {\bibinfo {title} {Pseudo-elasticity of a double network gel},\ }\href@noop {} {\bibfield  {journal} {\bibinfo  {journal} {Soft Matter}\ }\textbf {\bibinfo {volume} {7}},\ \bibinfo {pages} {8576} (\bibinfo {year} {2011})}\BibitemShut {NoStop}%
\bibitem [{\citenamefont {Ducrot}\ \emph {et~al.}(2015)\citenamefont {Ducrot}, \citenamefont {Montes},\ and\ \citenamefont {Creton}}]{ducrot2015structure}%
  \BibitemOpen
  \bibfield  {author} {\bibinfo {author} {\bibfnamefont {E.}~\bibnamefont {Ducrot}}, \bibinfo {author} {\bibfnamefont {H.}~\bibnamefont {Montes}},\ and\ \bibinfo {author} {\bibfnamefont {C.}~\bibnamefont {Creton}},\ }\bibfield  {title} {\bibinfo {title} {Structure of tough multiple network elastomers by small angle neutron scattering},\ }\href@noop {} {\bibfield  {journal} {\bibinfo  {journal} {Macromolecules}\ }\textbf {\bibinfo {volume} {48}},\ \bibinfo {pages} {7945} (\bibinfo {year} {2015})}\BibitemShut {NoStop}%
\bibitem [{\citenamefont {Huang}\ \emph {et~al.}(2007)\citenamefont {Huang}, \citenamefont {Furukawa}, \citenamefont {Tanaka}, \citenamefont {Nakajima}, \citenamefont {Osada},\ and\ \citenamefont {Gong}}]{huang2007importance}%
  \BibitemOpen
  \bibfield  {author} {\bibinfo {author} {\bibfnamefont {M.}~\bibnamefont {Huang}}, \bibinfo {author} {\bibfnamefont {H.}~\bibnamefont {Furukawa}}, \bibinfo {author} {\bibfnamefont {Y.}~\bibnamefont {Tanaka}}, \bibinfo {author} {\bibfnamefont {T.}~\bibnamefont {Nakajima}}, \bibinfo {author} {\bibfnamefont {Y.}~\bibnamefont {Osada}},\ and\ \bibinfo {author} {\bibfnamefont {J.~P.}\ \bibnamefont {Gong}},\ }\bibfield  {title} {\bibinfo {title} {Importance of entanglement between first and second components in high-strength double network gels},\ }\href@noop {} {\bibfield  {journal} {\bibinfo  {journal} {Macromolecules}\ }\textbf {\bibinfo {volume} {40}},\ \bibinfo {pages} {6658} (\bibinfo {year} {2007})}\BibitemShut {NoStop}%
\bibitem [{\citenamefont {Fukao}\ \emph {et~al.}(2020)\citenamefont {Fukao}, \citenamefont {Nakajima}, \citenamefont {Nonoyama}, \citenamefont {Kurokawa}, \citenamefont {Kawai},\ and\ \citenamefont {Gong}}]{fukao2020effect}%
  \BibitemOpen
  \bibfield  {author} {\bibinfo {author} {\bibfnamefont {K.}~\bibnamefont {Fukao}}, \bibinfo {author} {\bibfnamefont {T.}~\bibnamefont {Nakajima}}, \bibinfo {author} {\bibfnamefont {T.}~\bibnamefont {Nonoyama}}, \bibinfo {author} {\bibfnamefont {T.}~\bibnamefont {Kurokawa}}, \bibinfo {author} {\bibfnamefont {T.}~\bibnamefont {Kawai}},\ and\ \bibinfo {author} {\bibfnamefont {J.~P.}\ \bibnamefont {Gong}},\ }\bibfield  {title} {\bibinfo {title} {Effect of relative strength of two networks on the internal fracture process of double network hydrogels as revealed by in situ small-angle x-ray scattering},\ }\href@noop {} {\bibfield  {journal} {\bibinfo  {journal} {Macromolecules}\ }\textbf {\bibinfo {volume} {53}},\ \bibinfo {pages} {1154} (\bibinfo {year} {2020})}\BibitemShut {NoStop}%
\bibitem [{\citenamefont {Ju}\ \emph {et~al.}(2024)\citenamefont {Ju}, \citenamefont {Sanoja}, \citenamefont {Cipelletti}, \citenamefont {Ciccotti}, \citenamefont {Zhu}, \citenamefont {Narita}, \citenamefont {Yuen~Hui},\ and\ \citenamefont {Creton}}]{ju2024role}%
  \BibitemOpen
  \bibfield  {author} {\bibinfo {author} {\bibfnamefont {J.}~\bibnamefont {Ju}}, \bibinfo {author} {\bibfnamefont {G.~E.}\ \bibnamefont {Sanoja}}, \bibinfo {author} {\bibfnamefont {L.}~\bibnamefont {Cipelletti}}, \bibinfo {author} {\bibfnamefont {M.}~\bibnamefont {Ciccotti}}, \bibinfo {author} {\bibfnamefont {B.}~\bibnamefont {Zhu}}, \bibinfo {author} {\bibfnamefont {T.}~\bibnamefont {Narita}}, \bibinfo {author} {\bibfnamefont {C.}~\bibnamefont {Yuen~Hui}},\ and\ \bibinfo {author} {\bibfnamefont {C.}~\bibnamefont {Creton}},\ }\bibfield  {title} {\bibinfo {title} {Role of molecular damage in crack initiation mechanisms of tough elastomers},\ }\href@noop {} {\bibfield  {journal} {\bibinfo  {journal} {Proceedings of the National Academy of Sciences}\ }\textbf {\bibinfo {volume} {121}},\ \bibinfo {pages} {e2410515121} (\bibinfo {year} {2024})}\BibitemShut {NoStop}%
\bibitem [{\citenamefont {Divoux}\ \emph {et~al.}(2024)\citenamefont {Divoux}, \citenamefont {Agoritsas}, \citenamefont {Aime}, \citenamefont {Barentin}, \citenamefont {Barrat}, \citenamefont {Benzi}, \citenamefont {Berthier}, \citenamefont {Bi}, \citenamefont {Biroli}, \citenamefont {Bonn} \emph {et~al.}}]{divoux2024ductile}%
  \BibitemOpen
  \bibfield  {author} {\bibinfo {author} {\bibfnamefont {T.}~\bibnamefont {Divoux}}, \bibinfo {author} {\bibfnamefont {E.}~\bibnamefont {Agoritsas}}, \bibinfo {author} {\bibfnamefont {S.}~\bibnamefont {Aime}}, \bibinfo {author} {\bibfnamefont {C.}~\bibnamefont {Barentin}}, \bibinfo {author} {\bibfnamefont {J.-L.}\ \bibnamefont {Barrat}}, \bibinfo {author} {\bibfnamefont {R.}~\bibnamefont {Benzi}}, \bibinfo {author} {\bibfnamefont {L.}~\bibnamefont {Berthier}}, \bibinfo {author} {\bibfnamefont {D.}~\bibnamefont {Bi}}, \bibinfo {author} {\bibfnamefont {G.}~\bibnamefont {Biroli}}, \bibinfo {author} {\bibfnamefont {D.}~\bibnamefont {Bonn}}, \emph {et~al.},\ }\bibfield  {title} {\bibinfo {title} {Ductile-to-brittle transition and yielding in soft amorphous materials: perspectives and open questions},\ }\href@noop {} {\bibfield  {journal} {\bibinfo  {journal} {Soft Matter}\ }\textbf {\bibinfo {volume} {20}},\ \bibinfo {pages} {6868} (\bibinfo {year} {2024})}\BibitemShut {NoStop}%
\bibitem [{\citenamefont {Popovi{\'c}}\ \emph {et~al.}(2018)\citenamefont {Popovi{\'c}}, \citenamefont {de~Geus},\ and\ \citenamefont {Wyart}}]{popovic2018elastoplastic}%
  \BibitemOpen
  \bibfield  {author} {\bibinfo {author} {\bibfnamefont {M.}~\bibnamefont {Popovi{\'c}}}, \bibinfo {author} {\bibfnamefont {T.~W.}\ \bibnamefont {de~Geus}},\ and\ \bibinfo {author} {\bibfnamefont {M.}~\bibnamefont {Wyart}},\ }\bibfield  {title} {\bibinfo {title} {Elastoplastic description of sudden failure in athermal amorphous materials during quasistatic loading},\ }\href@noop {} {\bibfield  {journal} {\bibinfo  {journal} {Physical Review E}\ }\textbf {\bibinfo {volume} {98}},\ \bibinfo {pages} {040901} (\bibinfo {year} {2018})}\BibitemShut {NoStop}%
\bibitem [{\citenamefont {Barlow}\ \emph {et~al.}(2020)\citenamefont {Barlow}, \citenamefont {Cochran},\ and\ \citenamefont {Fielding}}]{barlow2020ductile}%
  \BibitemOpen
  \bibfield  {author} {\bibinfo {author} {\bibfnamefont {H.~J.}\ \bibnamefont {Barlow}}, \bibinfo {author} {\bibfnamefont {J.~O.}\ \bibnamefont {Cochran}},\ and\ \bibinfo {author} {\bibfnamefont {S.~M.}\ \bibnamefont {Fielding}},\ }\bibfield  {title} {\bibinfo {title} {Ductile and brittle yielding in thermal and athermal amorphous materials},\ }\href@noop {} {\bibfield  {journal} {\bibinfo  {journal} {Physical Review Letters}\ }\textbf {\bibinfo {volume} {125}},\ \bibinfo {pages} {168003} (\bibinfo {year} {2020})}\BibitemShut {NoStop}%
\bibitem [{\citenamefont {Rossi}\ \emph {et~al.}(2022)\citenamefont {Rossi}, \citenamefont {Biroli}, \citenamefont {Ozawa}, \citenamefont {Tarjus},\ and\ \citenamefont {Zamponi}}]{rossi2022finite}%
  \BibitemOpen
  \bibfield  {author} {\bibinfo {author} {\bibfnamefont {S.}~\bibnamefont {Rossi}}, \bibinfo {author} {\bibfnamefont {G.}~\bibnamefont {Biroli}}, \bibinfo {author} {\bibfnamefont {M.}~\bibnamefont {Ozawa}}, \bibinfo {author} {\bibfnamefont {G.}~\bibnamefont {Tarjus}},\ and\ \bibinfo {author} {\bibfnamefont {F.}~\bibnamefont {Zamponi}},\ }\bibfield  {title} {\bibinfo {title} {Finite-disorder critical point in the yielding transition of elastoplastic models},\ }\href@noop {} {\bibfield  {journal} {\bibinfo  {journal} {Physical Review Letters}\ }\textbf {\bibinfo {volume} {129}},\ \bibinfo {pages} {228002} (\bibinfo {year} {2022})}\BibitemShut {NoStop}%
\bibitem [{\citenamefont {Picard}\ \emph {et~al.}(2004)\citenamefont {Picard}, \citenamefont {Ajdari}, \citenamefont {Lequeux},\ and\ \citenamefont {Bocquet}}]{picard2004elastic}%
  \BibitemOpen
  \bibfield  {author} {\bibinfo {author} {\bibfnamefont {G.}~\bibnamefont {Picard}}, \bibinfo {author} {\bibfnamefont {A.}~\bibnamefont {Ajdari}}, \bibinfo {author} {\bibfnamefont {F.}~\bibnamefont {Lequeux}},\ and\ \bibinfo {author} {\bibfnamefont {L.}~\bibnamefont {Bocquet}},\ }\bibfield  {title} {\bibinfo {title} {Elastic consequences of a single plastic event: A step towards the microscopic modeling of the flow of yield stress fluids},\ }\href@noop {} {\bibfield  {journal} {\bibinfo  {journal} {The European Physical Journal E}\ }\textbf {\bibinfo {volume} {15}},\ \bibinfo {pages} {371} (\bibinfo {year} {2004})}\BibitemShut {NoStop}%
\bibitem [{\citenamefont {Reid}\ \emph {et~al.}(2018)\citenamefont {Reid}, \citenamefont {Pashine}, \citenamefont {Wozniak}, \citenamefont {Jaeger}, \citenamefont {Liu}, \citenamefont {Nagel},\ and\ \citenamefont {de~Pablo}}]{reid2018auxetic}%
  \BibitemOpen
  \bibfield  {author} {\bibinfo {author} {\bibfnamefont {D.~R.}\ \bibnamefont {Reid}}, \bibinfo {author} {\bibfnamefont {N.}~\bibnamefont {Pashine}}, \bibinfo {author} {\bibfnamefont {J.~M.}\ \bibnamefont {Wozniak}}, \bibinfo {author} {\bibfnamefont {H.~M.}\ \bibnamefont {Jaeger}}, \bibinfo {author} {\bibfnamefont {A.~J.}\ \bibnamefont {Liu}}, \bibinfo {author} {\bibfnamefont {S.~R.}\ \bibnamefont {Nagel}},\ and\ \bibinfo {author} {\bibfnamefont {J.~J.}\ \bibnamefont {de~Pablo}},\ }\bibfield  {title} {\bibinfo {title} {Auxetic metamaterials from disordered networks},\ }\href@noop {} {\bibfield  {journal} {\bibinfo  {journal} {Proceedings of the National Academy of Sciences}\ }\textbf {\bibinfo {volume} {115}},\ \bibinfo {pages} {E1384} (\bibinfo {year} {2018})}\BibitemShut {NoStop}%
\bibitem [{\citenamefont {Ahmed}\ \emph {et~al.}(2014)\citenamefont {Ahmed}, \citenamefont {Nakajima}, \citenamefont {Kurokawa}, \citenamefont {Haque},\ and\ \citenamefont {Gong}}]{ahmed2014brittle}%
  \BibitemOpen
  \bibfield  {author} {\bibinfo {author} {\bibfnamefont {S.}~\bibnamefont {Ahmed}}, \bibinfo {author} {\bibfnamefont {T.}~\bibnamefont {Nakajima}}, \bibinfo {author} {\bibfnamefont {T.}~\bibnamefont {Kurokawa}}, \bibinfo {author} {\bibfnamefont {M.~A.}\ \bibnamefont {Haque}},\ and\ \bibinfo {author} {\bibfnamefont {J.~P.}\ \bibnamefont {Gong}},\ }\bibfield  {title} {\bibinfo {title} {Brittle--ductile transition of double network hydrogels: Mechanical balance of two networks as the key factor},\ }\href@noop {} {\bibfield  {journal} {\bibinfo  {journal} {Polymer}\ }\textbf {\bibinfo {volume} {55}},\ \bibinfo {pages} {914} (\bibinfo {year} {2014})}\BibitemShut {NoStop}%
\bibitem [{\citenamefont {Irving}\ and\ \citenamefont {Kirkwood}(1950)}]{irving1950statistical}%
  \BibitemOpen
  \bibfield  {author} {\bibinfo {author} {\bibfnamefont {J.}~\bibnamefont {Irving}}\ and\ \bibinfo {author} {\bibfnamefont {J.~G.}\ \bibnamefont {Kirkwood}},\ }\bibfield  {title} {\bibinfo {title} {The statistical mechanical theory of transport processes. iv. the equations of hydrodynamics},\ }\href@noop {} {\bibfield  {journal} {\bibinfo  {journal} {The Journal of chemical physics}\ }\textbf {\bibinfo {volume} {18}},\ \bibinfo {pages} {817} (\bibinfo {year} {1950})}\BibitemShut {NoStop}%
\bibitem [{\citenamefont {Hardy}(1982)}]{hardy1982formulas}%
  \BibitemOpen
  \bibfield  {author} {\bibinfo {author} {\bibfnamefont {R.~J.}\ \bibnamefont {Hardy}},\ }\bibfield  {title} {\bibinfo {title} {Formulas for determining local properties in molecular-dynamics simulations: Shock waves},\ }\href@noop {} {\bibfield  {journal} {\bibinfo  {journal} {The Journal of Chemical Physics}\ }\textbf {\bibinfo {volume} {76}},\ \bibinfo {pages} {622} (\bibinfo {year} {1982})}\BibitemShut {NoStop}%
\bibitem [{\citenamefont {Dussi}\ \emph {et~al.}(2020)\citenamefont {Dussi}, \citenamefont {Tauber},\ and\ \citenamefont {Van Der~Gucht}}]{dussi2020athermal}%
  \BibitemOpen
  \bibfield  {author} {\bibinfo {author} {\bibfnamefont {S.}~\bibnamefont {Dussi}}, \bibinfo {author} {\bibfnamefont {J.}~\bibnamefont {Tauber}},\ and\ \bibinfo {author} {\bibfnamefont {J.}~\bibnamefont {Van Der~Gucht}},\ }\bibfield  {title} {\bibinfo {title} {Athermal fracture of elastic networks: How rigidity challenges the unavoidable size-induced brittleness},\ }\href@noop {} {\bibfield  {journal} {\bibinfo  {journal} {Physical Review Letters}\ }\textbf {\bibinfo {volume} {124}},\ \bibinfo {pages} {018002} (\bibinfo {year} {2020})}\BibitemShut {NoStop}%
\bibitem [{\citenamefont {Tauber}\ \emph {et~al.}(2020)\citenamefont {Tauber}, \citenamefont {Dussi},\ and\ \citenamefont {Van Der~Gucht}}]{tauber2020microscopic}%
  \BibitemOpen
  \bibfield  {author} {\bibinfo {author} {\bibfnamefont {J.}~\bibnamefont {Tauber}}, \bibinfo {author} {\bibfnamefont {S.}~\bibnamefont {Dussi}},\ and\ \bibinfo {author} {\bibfnamefont {J.}~\bibnamefont {Van Der~Gucht}},\ }\bibfield  {title} {\bibinfo {title} {Microscopic insights into the failure of elastic double networks},\ }\href@noop {} {\bibfield  {journal} {\bibinfo  {journal} {Physical Review Materials}\ }\textbf {\bibinfo {volume} {4}},\ \bibinfo {pages} {063603} (\bibinfo {year} {2020})}\BibitemShut {NoStop}%
\bibitem [{\citenamefont {Noguchi}\ and\ \citenamefont {Yukawa}(2024)}]{noguchi2024fracture}%
  \BibitemOpen
  \bibfield  {author} {\bibinfo {author} {\bibfnamefont {H.}~\bibnamefont {Noguchi}}\ and\ \bibinfo {author} {\bibfnamefont {S.}~\bibnamefont {Yukawa}},\ }\bibfield  {title} {\bibinfo {title} {Fracture process of composite materials in a spring network model},\ }\href@noop {} {\bibfield  {journal} {\bibinfo  {journal} {Physical Review E}\ }\textbf {\bibinfo {volume} {110}},\ \bibinfo {pages} {045001} (\bibinfo {year} {2024})}\BibitemShut {NoStop}%
\bibitem [{\citenamefont {Zhang}\ \emph {et~al.}(2018)\citenamefont {Zhang}, \citenamefont {Liu}, \citenamefont {Wang}, \citenamefont {Tang}, \citenamefont {Hu}, \citenamefont {Lu},\ and\ \citenamefont {Suo}}]{zhang2018fatigue}%
  \BibitemOpen
  \bibfield  {author} {\bibinfo {author} {\bibfnamefont {W.}~\bibnamefont {Zhang}}, \bibinfo {author} {\bibfnamefont {X.}~\bibnamefont {Liu}}, \bibinfo {author} {\bibfnamefont {J.}~\bibnamefont {Wang}}, \bibinfo {author} {\bibfnamefont {J.}~\bibnamefont {Tang}}, \bibinfo {author} {\bibfnamefont {J.}~\bibnamefont {Hu}}, \bibinfo {author} {\bibfnamefont {T.}~\bibnamefont {Lu}},\ and\ \bibinfo {author} {\bibfnamefont {Z.}~\bibnamefont {Suo}},\ }\bibfield  {title} {\bibinfo {title} {Fatigue of double-network hydrogels},\ }\href@noop {} {\bibfield  {journal} {\bibinfo  {journal} {Engineering Fracture Mechanics}\ }\textbf {\bibinfo {volume} {187}},\ \bibinfo {pages} {74} (\bibinfo {year} {2018})}\BibitemShut {NoStop}%
\bibitem [{\citenamefont {Chen}\ \emph {et~al.}(2016)\citenamefont {Chen}, \citenamefont {Yan}, \citenamefont {Zhu}, \citenamefont {Chen}, \citenamefont {Jiang}, \citenamefont {Wei}, \citenamefont {Huang}, \citenamefont {Yang}, \citenamefont {Liu},\ and\ \citenamefont {Zheng}}]{chen2016improvement}%
  \BibitemOpen
  \bibfield  {author} {\bibinfo {author} {\bibfnamefont {Q.}~\bibnamefont {Chen}}, \bibinfo {author} {\bibfnamefont {X.}~\bibnamefont {Yan}}, \bibinfo {author} {\bibfnamefont {L.}~\bibnamefont {Zhu}}, \bibinfo {author} {\bibfnamefont {H.}~\bibnamefont {Chen}}, \bibinfo {author} {\bibfnamefont {B.}~\bibnamefont {Jiang}}, \bibinfo {author} {\bibfnamefont {D.}~\bibnamefont {Wei}}, \bibinfo {author} {\bibfnamefont {L.}~\bibnamefont {Huang}}, \bibinfo {author} {\bibfnamefont {J.}~\bibnamefont {Yang}}, \bibinfo {author} {\bibfnamefont {B.}~\bibnamefont {Liu}},\ and\ \bibinfo {author} {\bibfnamefont {J.}~\bibnamefont {Zheng}},\ }\bibfield  {title} {\bibinfo {title} {Improvement of mechanical strength and fatigue resistance of double network hydrogels by ionic coordination interactions},\ }\href@noop {} {\bibfield  {journal} {\bibinfo  {journal} {Chemistry of Materials}\ }\textbf {\bibinfo {volume} {28}},\ \bibinfo {pages} {5710} (\bibinfo {year} {2016})}\BibitemShut {NoStop}%
\bibitem [{\citenamefont {Chen}\ \emph {et~al.}(2015)\citenamefont {Chen}, \citenamefont {Zhu}, \citenamefont {Chen}, \citenamefont {Yan}, \citenamefont {Huang}, \citenamefont {Yang},\ and\ \citenamefont {Zheng}}]{chen2015novel}%
  \BibitemOpen
  \bibfield  {author} {\bibinfo {author} {\bibfnamefont {Q.}~\bibnamefont {Chen}}, \bibinfo {author} {\bibfnamefont {L.}~\bibnamefont {Zhu}}, \bibinfo {author} {\bibfnamefont {H.}~\bibnamefont {Chen}}, \bibinfo {author} {\bibfnamefont {H.}~\bibnamefont {Yan}}, \bibinfo {author} {\bibfnamefont {L.}~\bibnamefont {Huang}}, \bibinfo {author} {\bibfnamefont {J.}~\bibnamefont {Yang}},\ and\ \bibinfo {author} {\bibfnamefont {J.}~\bibnamefont {Zheng}},\ }\bibfield  {title} {\bibinfo {title} {A novel design strategy for fully physically linked double network hydrogels with tough, fatigue resistant, and self-healing properties},\ }\href@noop {} {\bibfield  {journal} {\bibinfo  {journal} {Advanced Functional Materials}\ }\textbf {\bibinfo {volume} {25}},\ \bibinfo {pages} {1598} (\bibinfo {year} {2015})}\BibitemShut {NoStop}%
\end{thebibliography}

%

\section{End Matter}\label{end_matter}

\begin{figure}[!b]
\includegraphics[width=0.95\columnwidth]{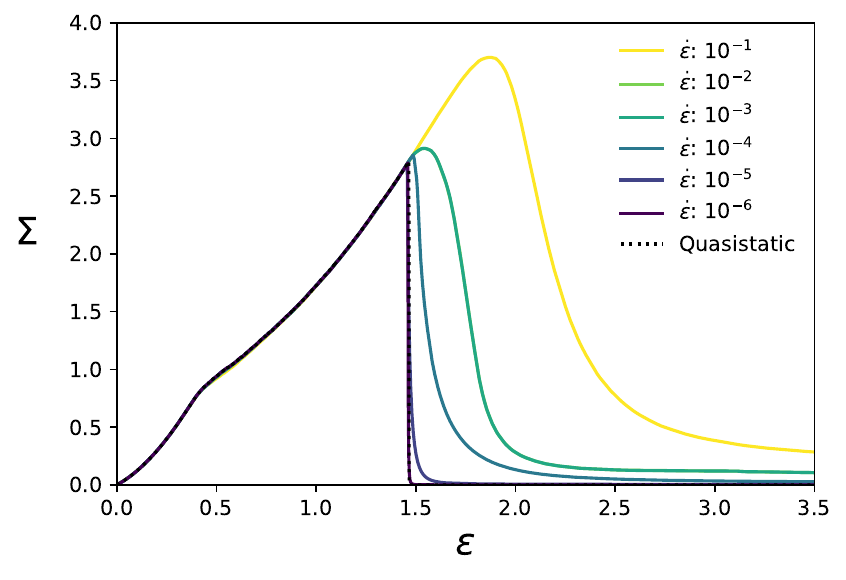}
\caption{Stress-strain curves for a double network stretched at a finite rate $\dot{\epsilon}>0$, showing also the approach to the quasistatic limit $\dot{\epsilon}\to 0$ explored in the main text. Parameters otherwise as in Fig.~\ref{fig:stressStrain}a).}
\label{fig:finiteRate}
\end{figure}

In the main text, we performed simulations using an algorithm that implements the limit of quasistatic shear. Finally, we show in Fig.~\ref{fig:finiteRate} that the introduction of a finite shear rate does not qualitatively change our key finding of a greatly improved toughness in a material with a double network structure.

\end{document}